\documentclass[draft]{agujournal2019}
\usepackage{url} 
\usepackage{lineno}
\usepackage[inline]{trackchanges} 
\usepackage{siunitx}
\usepackage{soul}
\usepackage{aas_macros}
\usepackage{natbib}
\usepackage{xcolor}

\draftfalse

\journalname{JGR: Atmospheres}

\begin{document}

%
%

\title{Atmospheric ionization rates during a geomagnetic reversal}

%
%




\authors{J. Svensmark\affil{1}}

\affiliation{1}{Atmospheric, Oceanic and Planetary Physics, Department of Physics, University of Oxford, Oxford, UK}

\correspondingauthor{Jacob Svensmark}{jacob.svensmark@physics.ox.ac.uk}




\begin{keypoints}
\item During the course of a geomagnetic reversal, global ionization rates are heightened.
\item The effect is most prominent at central latitudes, and increases with altitude.
\item The amplitude of the solar 11-year ionization modulation is highly dependent on the dipole strength.
\end{keypoints}

%
%

%
%


\begin{abstract}
The Matuyama-Brunhes reversal of Earth's magnetic dipole field took place 0.78 Ma ago, and detailed temporally resolved paleomagnetic data are available for this period. A geomagnetic reversal is expected to impact the cosmic ray flux, which in turn might impact atmospheric ionization rates. In this study a model that yields atmospheric ionization for the entire globe based on an input magnetic field is presented. 
Taking the time dependent paleomagnetic data as input, a 3D time series of the atmospheric ionization rates during the reversal is produced. We show, that as the dipole field weakens, the atmospheric ionization increases at low latitudes. The increase is ca.\ 25\% at the surface and up to a factor of 5 in the upper atmosphere. Globally, ionization rates increase around 13\% at the surface and up to a factor of 2 in the upper atmosphere, whereas polar regions are largely unaffected. Finally, the change in ionization due to the solar 11-year cycle is greatly affected by the reversal. The relative change in atmospheric ionization between solar-minimum and solar-maximum varies between 2 and two orders of magnitude. All atmospheric ionization data is made available for download.
\end{abstract}


\section{Introduction}
It has been hypothesized that galactic cosmic rays affect the terrestrial climate \citep{ney1959,dickinson1975,tinsley1991}. Several correlations between cosmic ray and climate proxies have been observed on a range of timescales from days \citep{svensmark2009,svensmark2016} and years \citep{svensmark1997} to millions of years \citep{shaviv2002,svensmark2006,svensmark2012,svensmark2022}. Although these correlations are plentiful and diverse, their robustness have been discussed. In particular, the lack of a generally accepted coupling mechanism has been a point of criticism. There are two main hypothesized channels through which cosmic rays may affect climate: The first channel couples cosmic rays to the global electric circuit, which in turn affects cloud droplet collision efficiencies and droplet freezing rates, altering cloud microphysics \citep{tinsley2000a,tinsley2000b}. The second channel concerns cloud condensation nuclei production, which may be affected by cosmic ray induced ion concentrations \citep{svensmark2007,kirkby2010}. The latter idea has lately been reinforced by laboratory experiments, in which the effect of a novel ion-induced aerosol condensation mechanism was observed \citep{svensmark2017}. If the correlations mentioned above originate from an underlying coupling between cosmic rays and climate, then this is something that requires to be further understood. It would imply, that any process which modulates the cosmic ray flux could result in a corresponding climate response. Therefore there is a need to quantify the ion concentration, distribution and temporal evolution during cosmic ray altering processes in Earth's atmosphere as part of further investigations. 

One such process is that of a geomagnetic reversal of Earth's dipole field. As it reverses, the magnetic field strength weakens, which allows for a higher number of cosmic rays to reach the top of Earth's atmosphere. Interestingly, geomagnetic reversals have previously been associated with climate changes \citep{kitaba2017}. In the present work, we set out to estimate the change in atmospheric ionization rates during the Matuyama-Brunhes magnetic reversal event at 0.78$\,$Ma BP. 


In a model similar to \cite{usoskin2006}, an spectral energy distribution of cosmic rays from the interstellar medium is modulated by the solar magnetic field. Then, as the cosmic rays progress toward Earth, its 3-dimensional magnetic field imposes further spatial cosmic ray distribution around the globe. Finally, when the remaining cosmic rays reach Earth's upper atmosphere, they initiate a cascade of secondary particles, which interact with atmospheric constituents and maintain a vertical atmospheric ionization profile. Since this model accounts for the variations in Earth's magnetic field during the time of the Matuyama-Brunhes reversal, the resulting ion-pair production rate $q(\lambda,\phi,h,t)$ becomes a function of longitude $\lambda$, latitude $\phi$, atmospheric height $h$ and time $t$. We show that the ionization rates change at a scale comparable to the solar 11-year cycle but in a completely different spatial pattern. Due to the prevalence of the geomagnetic dipole field component, the reversal increases ionization rates mainly in the equatorial regions. Furthermore, we show that the amount ionization, caused by the solar 11-year cycle between minimum and maximum is highly dependent on the geomagnetic field, which may alter its amplitude up to two orders of magnitude.  

The paper is organized as follows: In Section \ref{sec:theory}, we present the equation for the atmospheric ionization rate and details on the sources of its main constituents namely the cosmic ray flux spectrum (Section \ref{sec:gcr_spectrum}), the geomagnetic cutoff rigidity (Section \ref{sec:GEO_SMART}), and the ionization yield function for cosmic rays (\ref{sec:yield_function}). Under results Section \ref{sec:results}, the model is run for the present magnetic field, and compared to atmospheric observations of the ionization rate in Section \ref{sec:present_day_observations}. In Section \ref{sec:cutoff_rigidity_results} the geomagnetic cutoff rigidity is calculated using the paleomagnetic field data at all available times. The resulting spatial and temporal changes in atmospheric ionization are presented in section \ref{sec:ionization_during_reversal}, and the consequences for the solar 11-year ionization cycle is shown in Section \ref{sec:comparisons_with_sun}. Finally, the results are discussed and our findings concluded on in Section \ref{sec:discussion_conclusion}. The reader primarily interested in the resulting ion-distribution may go directly to Figure \ref{fig:ionization_profiles} and \ref{fig:ionization_temporal}, in which we display the latitudinal and temporal variations in ion production rates. The reader concerned with how the influence of the sun's 11-year cycle on ionization changes with the geomagnetic field may instead go directly to Figure \ref{fig:solaramp_vs_d}.

\section{Cosmic ray spectra, yield functions and ionization}\label{sec:theory}
The ionization rate $q$ throughout Earth's atmosphere at a given time $t$ can be determined by three components: 1) The cosmic ray flux spectrum as a function of kinetic energy $J(T)=dF/dT$ at 1$\,$AU. 2) The modulation of the cosmic ray spectrum by Earth's magnetic field at a given time, longitude $\lambda$ and latitude $\varphi$, as represented by the geomagnetic cutoff rigidity $R_\mathrm{c}(t,\lambda,\varphi)$. 3) The yield function $Y_A(R,h)$ describing the number ion-pairs an initial cosmic ray of species $A$ and rigidity $R$ will produce at height $h$ the atmosphere. From these, the ion-pair production rate $q$ can be found as 
\begin{equation}\label{eq:ionization}
    q(\lambda,\phi,h,t) =\sum _{\mathrm{A}} \int _{T_{\mathrm{c},\mathrm{A}}(t,\lambda,\phi)}^{\infty}Y_{\mathrm{A}}(T,h)J_{\mathrm{A}}(T)\,dT,
\end{equation}
where and $T_{\mathrm{c},\mathrm{A}}$ is the cutoff energy of cosmic ray particles of species $\mathrm{A}$ determined by the cutoff rigidity $R_\mathrm{c}$. We now describe how each of these components are obtained.

\subsection{The cosmic ray particle spectrum and solar modulation}\label{sec:gcr_spectrum}

\begin{figure}\hspace{-1.5cm}\center
\noindent\includegraphics[width=1\textwidth]{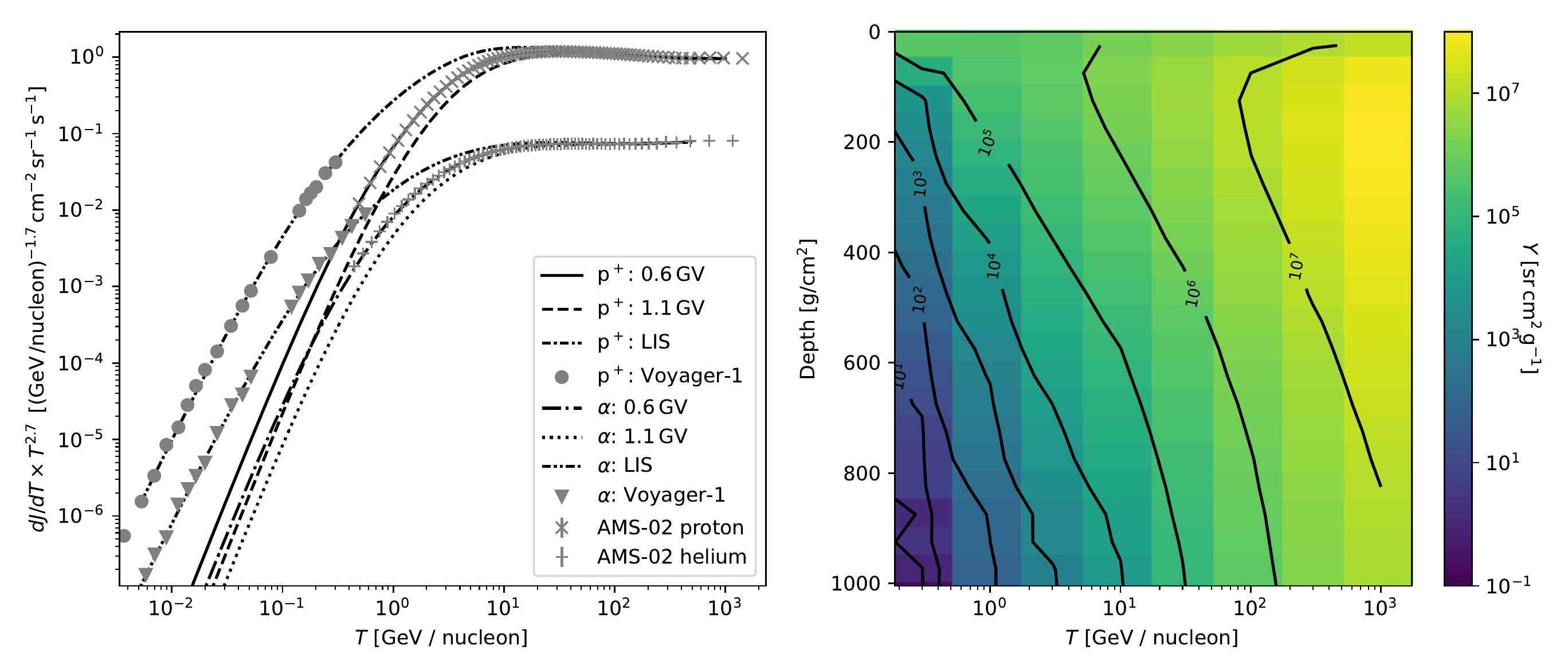}
\caption{Left: Differential cosmic ray energy spectra of protons and alpha particles at the top of the atmosphere using the forcefield approximation, with $\phi=600\,$MV (solar minimum) and $\phi=1100\,$MV (solar maximum) \cite{corti2016}. Also depicted are the AMS-02 measurements that the spectra are generated from. Right: Cosmic ray yield function for protons from Table 1 of \cite{usoskin2006}.}
\label{fig:spectra_and_yield}
\end{figure}

Measurements of the Local Interstellar Spectrum (LIS) of cosmic ray are available from the Voyager 1 satellite positioned just outside the heliosphere \citep{gurnett2013,stone2013}. As cosmic rays travel into the heliosphere, the LIS spectrum gets modulated by solar magnetic activity carried by the solar wind. A simple for describing solar modulation of cosmic rays comes in the form of the force-field approximation, which modifies the LIS based on a single parameter $\phi(t)$, assuming spherical symmetry and zero streaming. According to this, the differential cosmic ray flux becomes
\begin{equation}
\frac{dF}{dT}\left(T,\Phi\left(t\right)\right)=\frac{T\left(T+2M\right)}{\left(T+\Phi\left(t\right)\right)\left(T+\Phi\left(t\right)+2M\right)}\frac{dF_{\mathrm{LIS}}(T+\Phi(t))}{dT}.
\end{equation}
Here, $T$ is the kinetic energy of a nucleus, $M$ is its mass, $Z$ its charge and $\Phi(t)=Ze\phi(t)$. $\phi(t)$ is the modulation potential with units of electrical potential. The differential LIS spectrum $dF_{\mathrm{LIS}}/dT$ is here based on a fit of measurements of the LIS by the \textit{Voyager I} spacecraft, as well as measurements of the solar modulated spectrum inside the heliosphere, as measured by the AMS-02 instrument aboard the International Space Station:
\begin{eqnarray}\label{eq:LIS}
    \frac{dF_{LIS}}{dR}(R) = 
    &N_{0}&\left(
    1+\exp\left(\frac{\ln R-\mu}{\sigma}\right)
    \right)^{-1/\nu}R^{\gamma_1}\\
    &\times&\left(
    1+
    \left(
    \frac{R}{R_{b1}}\left(
    1+\left(\frac{R}{R_{b2}}\right)^{\Delta\gamma_2/s_2}
    \right)^{s_2}
    \right)^{\Delta\gamma_1/s_1}
    \right)^{s_1}.
\end{eqnarray}
All of the parameters are given in Table \ref{tab:LISfit} for the proton and helium component \citep{corti2016}. Note that this is formulated in terms of rigidity $R$, and needs to be rewritten via the chain rule to fit Equation \ref{eq:LIS}. With this in hand, we can represent the differential cosmic ray spectrum at different $\phi$ i.e. at different times during the solar cycle. Both the LIS and heliospheric spectra can be seen in Figure \ref{fig:spectra_and_yield} for the proton and helium species for modulation parameters written in the Figure caption. Furthermore, data from Voyager-1 and the AMS-02 instrument are shown for comparison. Note that all values have been multiplied by $T^{2.7}$ to flatten the slope of the upper end of the spectra. \cite{corti2016} uses \cite{adriani2013} for values of the solar modulation parameter. However, these values only span a time interval shorter than a full solar cycle. The solar modulation parameter has been calculated by \cite{usoskin2011} for 75 years of data using other versions of the force-field approximation. For this reason, we can instead extrapolate a linear relationship between the solar modulation parameter $\phi_{\mathrm{Usoskin}}(t)$ by \cite{usoskin2011} and $\phi_{\mathrm{Adriani}}(t)$ by \cite{adriani2013}, to obtain $\phi$ anywhere within a solar cycle. By fit, we get
\begin{equation}\label{eq:phis}
    \phi_{\mathrm{Adriani}}(t) = 0.9858954 \times \phi_{\mathrm{Usoskin}}(t)  + 41.6277179\,\mathrm{MV},
\end{equation}
where the $\phi$'s are in units of MV. From Equation \ref{eq:phis} we can obtain $\phi_{\mathrm{Adriani}}(t)$ anywhere in a solar cycle. In the following, $\phi$ we will be referring to values of $\phi_{\mathrm{Adriani}}(t)$. Furthermore, unless otherwise stated, we define $\phi=300\,$MV as the solar minimum value and $\phi=1100\,$MV as the solar maximum value.

 \begin{table}\label{tab:LISfit}
 \caption{Values used for the differential cosmic ray LIS of Equation \ref{eq:LIS} from \cite{corti2016}.}
 \centering
 \begin{tabular}{l r r}
 \hline
  Parameter  & Proton & Helium  \\
 \hline
   $N_0$ [m$^{-2}\,$sr$^{-1}\,$s$^{-1}\,$GV$^{-1}$] & $\num{11740}$ & $\num{11220}$  \\
   $\mu$             & $\num{-0.559}$  & $\num{0.197}$   \\
   $\sigma$          & $\num{0.563}$   & $\num{0.477}$   \\
   $\nu$             & $\num{0.431}$   & $\num{0.559}$   \\
   $\gamma_1$        & $\num{-2.4482}$ & $\num{-1.9017}$ \\
   $R_{b1}$ [GV]     & $\num{6.2}$     & $\num{2.14}$    \\
   $\Delta\gamma_1$  & $\num{-0.4227}$ & $\num{-1.1958}$ \\
   $s_1$             & $\num{-0.108}$  & $\num{-3.08}$   \\
   $R_{b2}$ [GV]     & $\num{545}$     & $\num{189}$     \\
   $\Delta\gamma_2$  & $\num{-0.6}$    & $\num{-0.332}$  \\
   $s_2$             & $\num{-0.4}$    & $\num{-0.232}$  \\
 \hline
 \end{tabular}
 \end{table}

\subsection{Earths magnetic field and reversal data and geomagnetic cutoff rigidity}\label{sec:GEO_SMART}
Geomagnetic reversal events are relatively short-lived. The most recent Matuyama-Brunhes magnetic reversal event at 0.78$\,$Ma BP had a duration of 28$\,$kyr \citep{clement2004,bogue2005}. Based on a compilation of high quality sedimentary and lava flow records, \citet{leonhardt2007} constructed a temporal spherical harmonic expansion of the magnetic field using a Bayesian inversion method. To understand the geomagnetic data, we must first introduce a theoretical basis for the geomagnetic field.

The geomagnetic field $\mathbf{B}_{\mathrm{int}}=-\nabla V_{\mathrm{int}}$ from internal sources at the boundary of Earth $a$ at time $t$ and location ($\theta$,$\phi$) can be described as a multi-pole spherical harmonic expansion to degree $L$ of its potential as:
\begin{equation}\label{eq:magnetic_potential}
V_{\mathrm{int}}=a\sum_{n=1}^L\sum_{m=0}^{n}(g_n^m\cos m\phi + h_n^m\sin m\phi)\left(\frac{a}{r}\right)^{n+1}P_n^m(\cos \phi).
\end{equation}
Here, $g_n^m$ and $h_n^m$ are the Gauss coefficients of the internal sources taken at Earth's surface, and $P_n^m$ are the associated Schmidt semi-normalized Legendre functions \citep{olsen2012}. The Gauss coefficients $g_n^m$ and $h_n^m$ determine the actual shape of the field, and for a variable field, they are functions of time. The Gauss coefficient's values depend on the radius at which they are calculated. If they are not provided at e.g. Earth's radius $a$ but at some other radius $r^{\prime}$ as $\{{g^{\prime}}_n^m(t),{h^{\prime}}_n^mc(t)\}$, they can be re-scaled to Earths surface, e.g. for use in Equation \ref{eq:magnetic_potential}, as
\begin{equation}
    g_n^m(t) = \left(  \frac{r^{\prime}}{a}\right)^{n+2} {g^{\prime}}_n^m(t),
\end{equation}
and 
\begin{equation}
    h_n^m(t) = \left( \frac{r^{\prime}}{a} \right)^{n+2} {h^{\prime}}_n^m(t).
\end{equation}
At degree $L$, there are $L(L+2)$ non-zero components representing the shape and intensity of the field. While these provide the complexity and extent of the geomagnetic field, it can be helpful to consider a simplified quantity that summarizes the field behaviour over time. One way to do this is by considering the energy of the magnetic field from its individual multipole components. The $n$'th component of the Mauersberger–Lowes spectrum $R_n$ shows the energy of the magnetic field due to the $n$'th multipole as:
\begin{equation}
R_n = \left(\frac{r^{\prime}}{a}\right)^{2n+4}(1+n)\sum _{m=0}^{n}\left[{g^{\prime}}_n^m(t)^2+{g^{\prime}}_n^m(t)^2\right].
\end{equation}
Here we will simply refer to it as ``the energy of n'th multipole''. 

This study's primary focus is on Earth's varying magnetic field's effect on atmospheric ionization rates. The geomagnetic field affects cosmic ray trajectories far beyond the extent of the terrestrial atmosphere. 
The cosmic ray particles follow complicated trajectories that depend on mass, charge, energy, position, and direction in the geomagnetic field. However, not all particles will arrive at the top of Earth's atmosphere and contribute to atmospheric ionization. A standard approach to this problem is to solve the reverse trajectory. Instead of starting the particle outside Earth's geomagnetic field, the trajectory is calculated backwards in time from the top of Earth's atmosphere by reversing the velocity vector and changing the sign of the particle's charge. For a given latitude and longitude, charged test particles of varying rigidity and directions are launched from the top of the Earth's atmosphere (20 km), and their trajectories are calculated numerically. If a particle of a given rigidity reaches a large enough distance from Earth, the rigidity is registered as 'allowed'. However, if a particle has a rigidity, so its trajectory intersects Earth's surface again, it is registered as 'disallowed'. Due to chaotic trajectories, there is not, in general, a sharp cutoff rigidity value but a band of rigidity values where trajectories are allowed and not allowed, called the penumbra. Therefore, the cutoff rigidity is defined as the midpoint between the highest disallowed rigidity and the lowest allowed rigidity. The \texttt{GEO\-MAG\-NE\-TIC CUT\-OFF \-RIGIDITY \-COM\-PU\-TER \-PRO\-GRAM} by Don F. Smart and Margaret A. Shea simulates precisely this process \citep{smart2000} for Earth's current magnetic field to the 11'th order. We modified the numerical code to take a variable magnetic field as input to calculate the cutoff rigidity at several time steps during the geomagnetic reversal. 

Leonhardt and Fabian provide the Gauss coefficients $\{{g^{\prime}}_n^m(t),{h^{\prime}}_n^mc(t)\}$ at the core-mantle boundary at radius $r^{\prime}=3480\,$km at 319 different times between 794.48$\,$kyr 763.77$\,$kyr before present. After scaling, the Gauss coefficients can then be directly substituted into the \texttt{GEOMAGNETIC CUTOFF RIGIDITY COM\-PU\-TER PRO\-GRAM} in place of the current magnetic field coefficients as it relies on Equation \ref{eq:magnetic_potential} for its calculations. This then outputs the cutoff rigidity $R_c(\lambda,\varphi,t)$ as function of longitude $\lambda$, latitude $\varphi$ and time $t$. Any given cutoff rigidity can be converted to a cutoff kinetic energy per nucleon $T_{c,A}$ for a given particle species $\mathrm{A}$, and inserted into the lower bound of the integral of Equation \ref{eq:ionization} to calculate the ionization rate.

\subsection{Cosmic ray showers using CORSIKA}\label{sec:yield_function}
When a primary cosmic ray particle collides with atmospheric atoms at the top of the atmosphere, it produces a cascade of secondary cosmic ray particles. These particles deposit part of their energy by ionization of atmospheric constituents as the shower propagates down towards the ground. A widely used numerical code that can simulate such events is the CORSIKA (COsmic Ray SImulations for KAscade) code. \citet{usoskin2006} uses CORSIKA to produce statistical ensembles of showers and energy deposits from primary cosmic rays at energy $T$ between 0.1$\,$GeV/nucleon and 1000$\,$GeV/nucleon at an incoming angle uniformly chosen between 0$^{\circ}$ and 90$^{\circ}$. The average energy deposited by the shower at a given height for the given initial kinetic energy then precisely makes up the yield function. In Tables 1 and 2 of \cite{usoskin2006}, the yield function has been calculated for protons and alpha particles, respectively, and in the right-hand panel of Figure \ref{fig:spectra_and_yield} on the right, we show the yield function for protons. With a yield function for protons and helium, a time-dependent cutoff rigidity map and a cosmic ray spectrum, we calculate ionization rates for the entire globe for both the present strength of the magnetic field and the strength of the magnetic field during the reversal. These results are presented in the next section.

\section{Results}\label{sec:results}
The cosmic ray spectra, yield functions and cutoff rigidity maps presented in the previous sections make it possible to calculate spatial ionization values for the entire globe. To see that it works, we first calculate present-day ionization rates. Therefore, we use the current magnetic field to produce a contemporary cutoff rigidity map to calculate ionization profiles and compare these to observational measurements performed throughout the solar cycle. 
Then as the model is confirmed to work for contemporary parameters, we calculate ionization rates for the reversal event. 

\subsection{Comparison with present day observations}\label{sec:present_day_observations}
\begin{figure}\hspace{-1.5cm}\center
\noindent\includegraphics[width=1\textwidth]{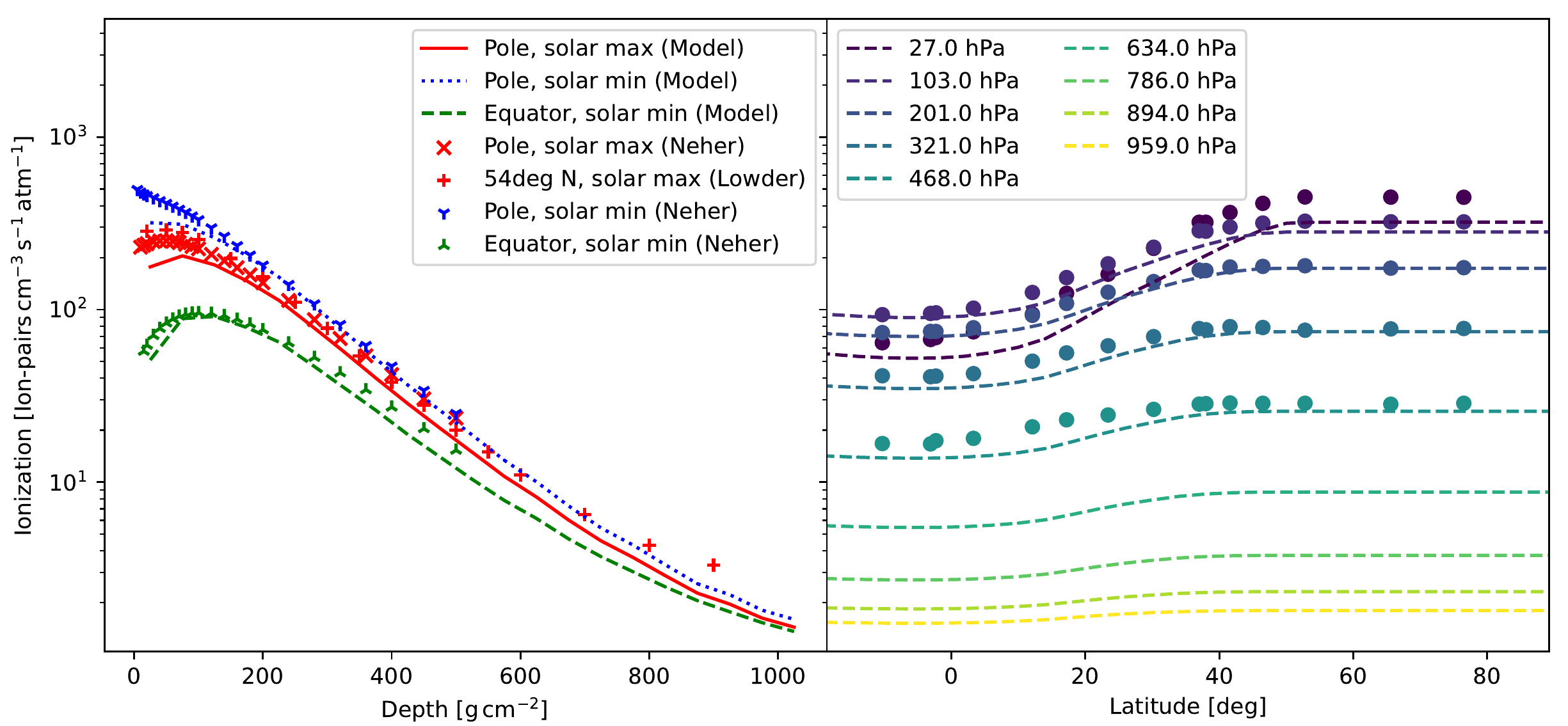}
\caption{\textit{Left:} Vertical ionization profiles, calculated by the model presented in this paper for a present day magnetic field, compared to similar observations. The green dashed line shows the ionization during solar minimum ($\phi=455\,$MV) at the Equator and 80$^{\circ}\,$W. The blue dotted line shows the polar profile at solar minimum ($\phi=445\,$MV) for the model, and the red line shows the same but for solar maximum ($\phi=1.025$\,MV). The ditto colored symbols show observations from \cite{neher1967,neher1971,lowder1972} at similar locations performed at times corresponding to the $\phi$'s used in the model. \textit{Right}: Latitude profiles for the model (lines) at solar minimum ($\phi=455\,$MV) averaged between 80$^{\circ}\,$W and 70$^{\circ}\,$W, compared to observations from \cite{neher1967} spanning similar longitudes.}
\label{fig:present_day_ionization}
\end{figure}

The left panel of Figure \ref{fig:present_day_ionization} shows the vertical profiles of the modelled ionization rate using a contemporary magnetic field for three different times and places. The green dashed line shows the atmospheric ionization profile for the solar minimum year of 1965 in May at the Equator at 80$^{\circ}\,$W. The blue dotted line is the same but taken at the North Pole. Finally, the solid red line shows the North Pole profile at the solar maximum year of 1958. Corresponding measurements are shown as the ``tri-up'', ``tri-down'' and ``plus'' symbols respectively \citep{neher1967,neher1971,lowder1972}. On the right side of Figure \ref{fig:present_day_ionization} latitudinal ionization profiles are shown for 9 different atmospheric pressures averaged between 80$^{\circ}\,$W and 70$^{\circ}\,$W. For the five upper pressures, latitudinal observations are shown in similar colours \citep{neher1967}. Generally, the model shows a slightly low but overall good correspondence to the observations and, in essence, reproduces the earlier work on the contemporary ionization distribution by \cite{usoskin2006}.

\subsection{Cutoff rigidity during the reversal}\label{sec:cutoff_rigidity_results}
\begin{figure}
\noindent\includegraphics[width=1.05\textwidth]{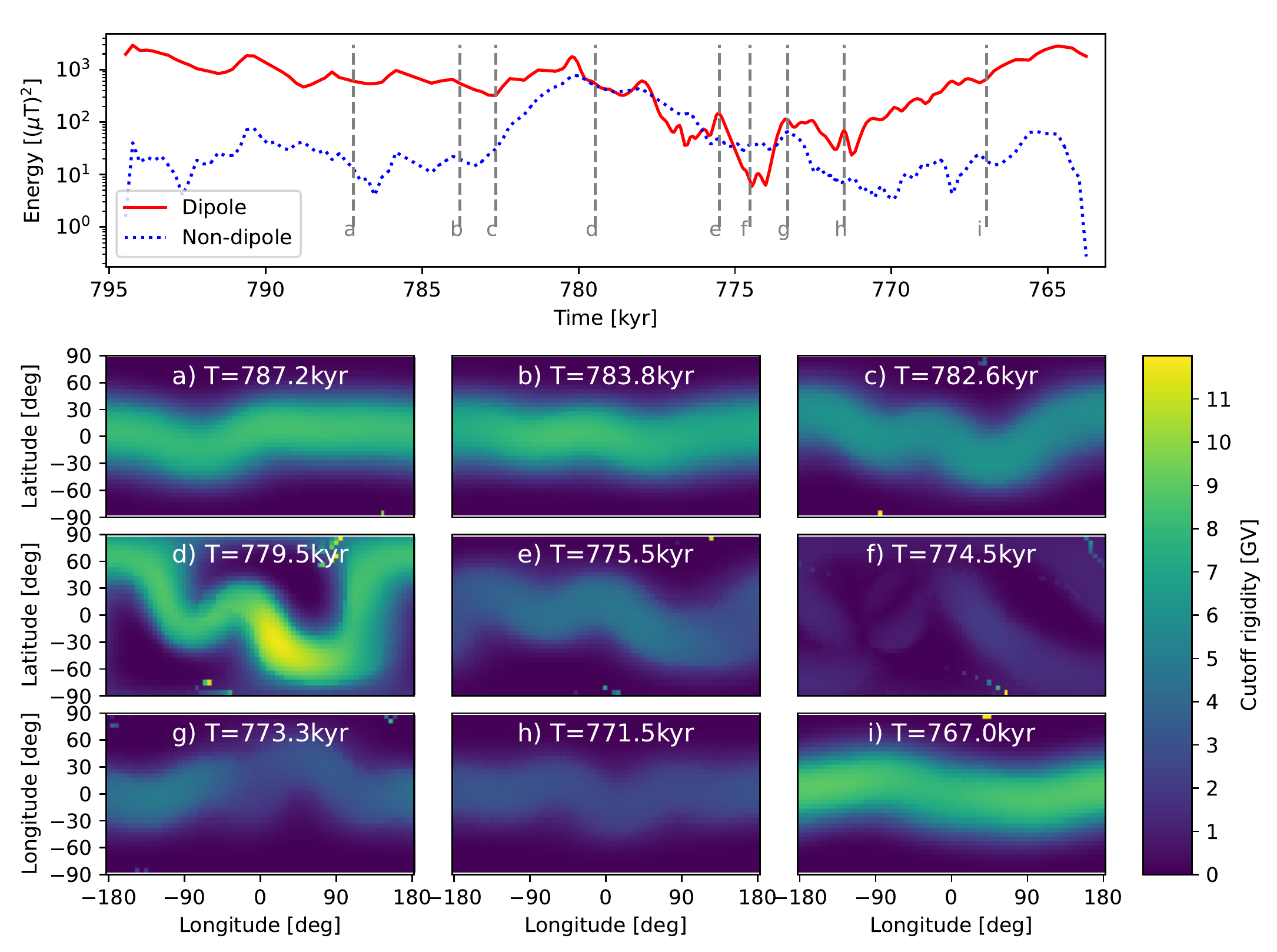}\centering
\caption{\textit{Upper panel:} Energy of the magnetic field from the dipole component (solid red) and the sum of higher order multipoles i.e. the 4-, 8- and 16-pole components (blue dashed) during the Matuyama-Brunhes reversal, as quantified by the $R_n$ of the Mauersberger–Lowes spectrum. This is essentially a reproduction of Figure 8 of \cite{leonhardt2007}, including the highlighted times a) through i). \textit{Lower panels:} Maps of calculated cutoff rigidities for the 9 different time steps during the Matuyama-Brunhes reversal event marked in the upper panel, using the magnetic field information from \cite{leonhardt2007}}.
\label{fig:cutoff_rigidity_reversal}
\end{figure}

With confidence in the ability of our model to calculate realistic ionization functions for a given magnetic field and thus cutoff rigidity map, we proceed towards calculating historic ionization fields using the geomagnetic field data recorded for the Matuyama-Brunhes reversal. First, the Mauersberger–Lowes spectral components are consulted to understand better how the field varies during this event. The top panel of Figure \ref{fig:cutoff_rigidity_reversal} shows the Mauersberger–Lowes dipole and multipole (sum of the 4-, 8-, and 16-pole) component of the geomagnetic field during the reversal. At the offset of the data set, the dipole component is dominant with up to two orders of magnitude. Onwards from point b) the field flares up, and the multipole components become comparable in magnitude with the dipole. Then, from point d), the total field decreases until point f) from where the dipole component grows back and the multipoles continue to drop towards pre-reversal magnitudes. While it is not directly visible through this figure, the dipole field has now reversed its orientation. 

By running the \texttt{GEO\-MAG\-NE\-TIC CUT\-OFF \-RI\-GI\-DI\-TY \-COM\-PU\-TER \-PRO\-GRAM} introduced in section \ref{sec:GEO_SMART} using the Gauss coefficients summarized in the top of Figure \ref{fig:cutoff_rigidity_reversal}, we arrive at a series of cutoff rigidity maps for the entire globe. In the bottom part of Figure \ref{fig:cutoff_rigidity_reversal}, such maps can be seen for each of the nine times marked in the upper panel. The panes reflect many of the same features that are mentioned for the Mauersberger–Lowes spectrum above: Initially, Earth's magnetic field has a higher cutoff rigidity around the equator due to the pronounced dipole structure of the geomagnetic field. Conversely, the cutoff is near 0$\,$GV in the polar regions, which is in line with the present-day magnetic field and cutoff rigidity calculations \citep{smart2005}. Then, as the reversal progresses, the equatorial dipole belt of high cutoff rigidity becomes more curly as the multipole components grow towards point d). Then as the field weakens towards point f), the cutoff rigidity reduces to near 0$\,$GV for the entire globe. At panel f), the dipole component is at its lowest, and the cutoff rigidity map is near 0$\,$GV globally. Higher-order terms of the magnetic field dominate the remaining structure of the cutoff rigidity map. From panel, g), the dipole component of the magnetic field increases and so does the dipole structure of the cutoff rigidity distribution, resulting in higher cutoffs at mid-latitudes. Through panels h) and i), the dipole component grows. In contrast, the multipole components decrease, and the cutoff rigidity shows a clear dipole structure with a latitudinal cutoff profile, similar to panel a). Below, we shall mention these times in the same manner a), b), c) etc., and be referring to the times indicated in Figure \ref{fig:cutoff_rigidity_reversal}. 

\subsection{Ionization during the reversal}\label{sec:ionization_during_reversal}
\begin{figure}\hspace{-1.5cm}\centering
\noindent\includegraphics[width=1.\textwidth]{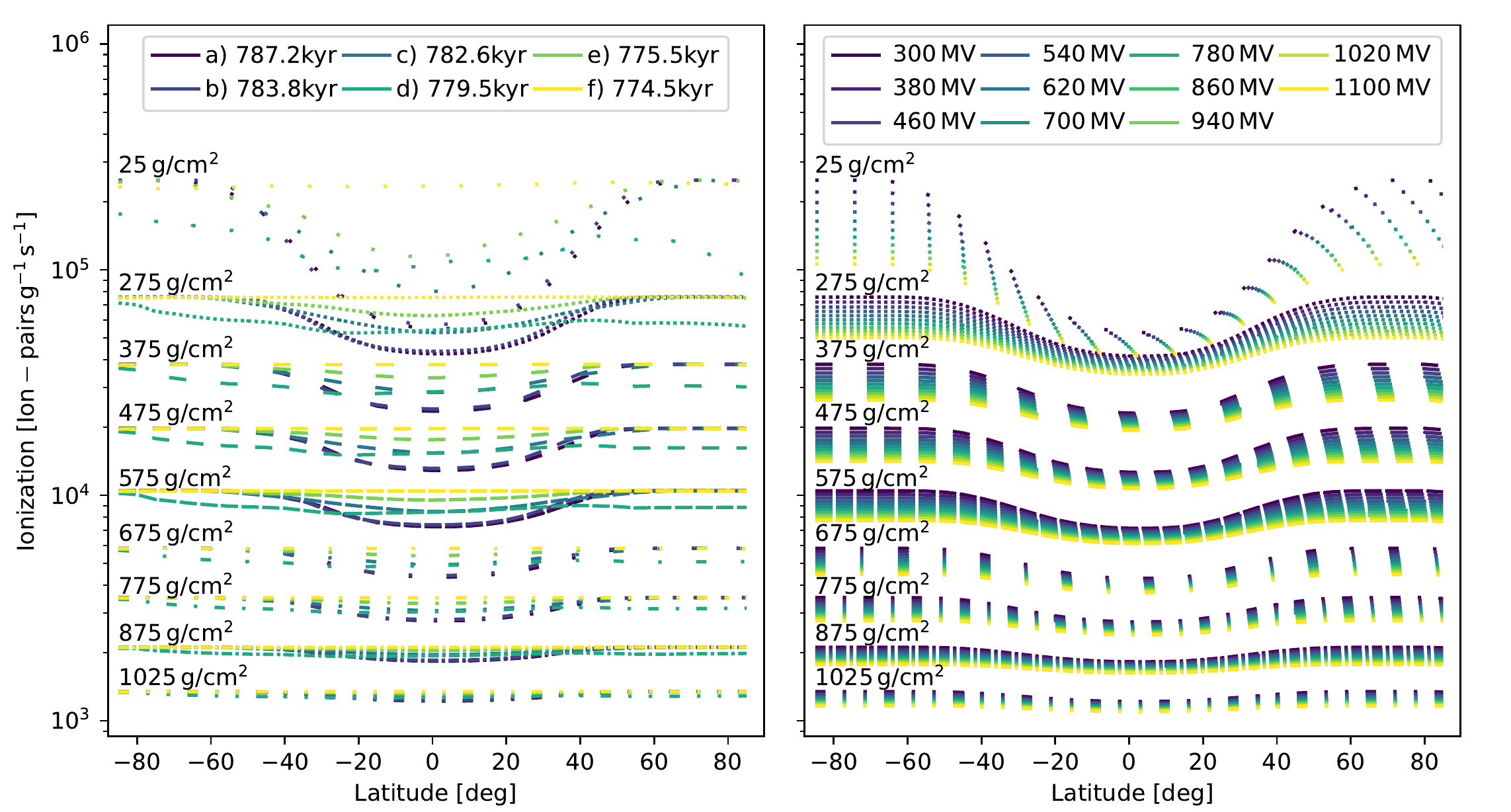}
\caption{Left: Ion production rate at 9 different pressure surfaces for the first 6 points a) to f) in time during the Matuyama-Brunhes reversal event, at solar minimum $\phi=300\,$MV. Right: The same as on the left, but instead for 11 different values of the solar modulation parameter, spanning from solar minimum to solar maximum, for the time $T=767.0\,$kyr.}
\label{fig:ionization_profiles}
\end{figure}

\begin{figure}\hspace{-1.5cm}\centering
\noindent\includegraphics[width=1.0\textwidth]{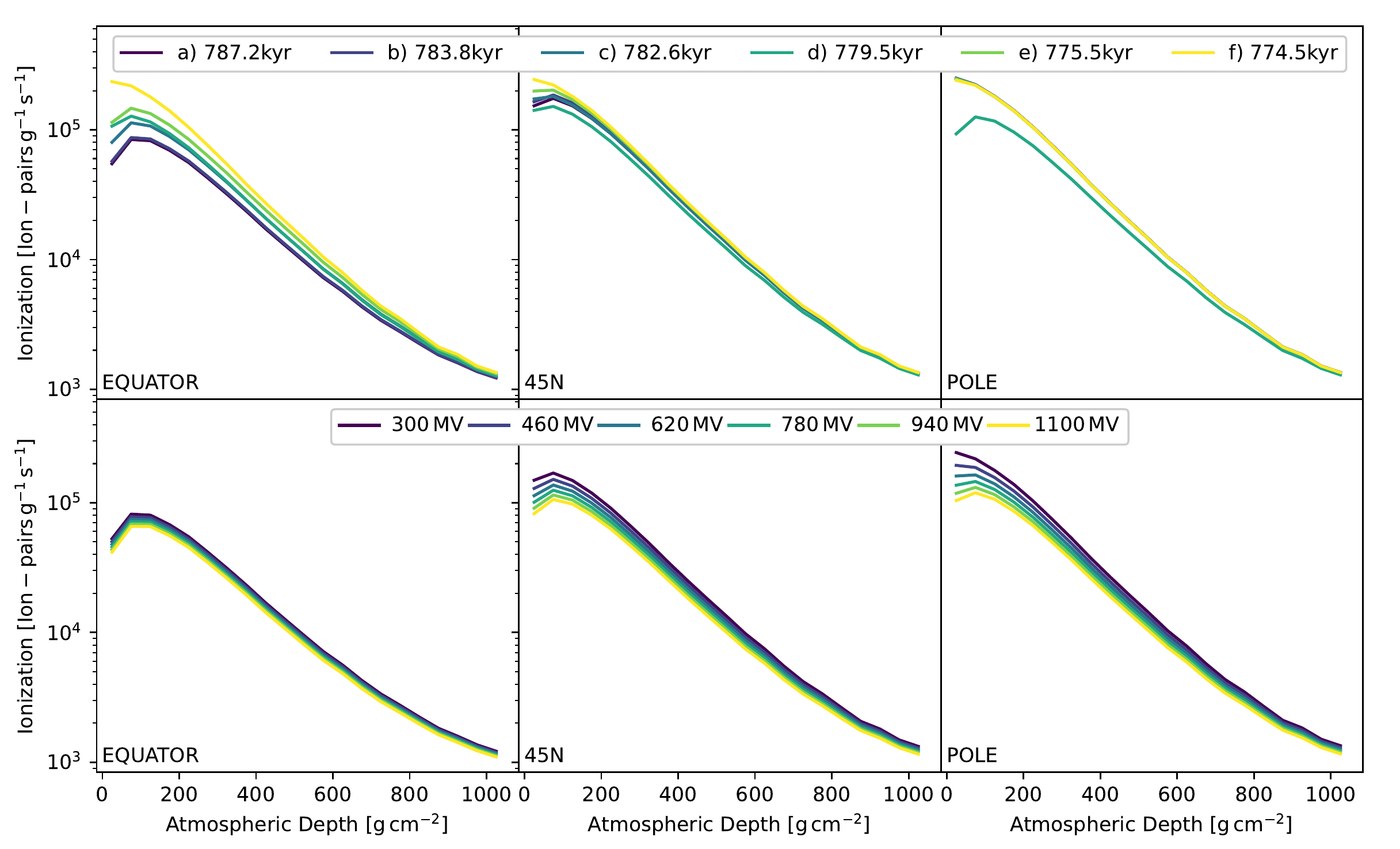} 
\caption{Ion production rate at the equator (left), 45 degrees latitude North (middle) and at the North Pole (right) for the 6 different times during the Matuyama-Brunhes reversal event from point a) to f) (see Figure \ref{fig:cutoff_rigidity_reversal}) at solar minimum $\phi=300\,$MV (upper panels) and 6 different values of the solar modulation parameter, spanning from solar minimum to solar maximum, for the time $T=767.0\,$kyr (lower panels).}
\label{fig:ionization_profiles_vertical}
\end{figure}

The magnetic field reversal and its consequences for the geomagnetic cutoff rigidity provides the temporal dimension of the ionization rate $q(\lambda,\phi,h,t)$ of Equation \ref{eq:ionization}. The left panel of Figure \ref{fig:ionization_profiles} shows latitudinal ionization profiles for the points a) to f) in time during the Matuyama-Brunhes reversal. At the time, a) the field's clear and pronounced dipole structure is visible as a lower ionization rate around central latitudes and a higher rate at the North and South Pole. The dipole energy decreases from a) to c), leading to an increased ionization rate at central latitudes. Then, the ionization rate decreases globally as the dipole and higher-order multipole components vary towards equal magnitudes at point d). Towards point e), the dipole and higher-order multipole Energy decreases more than an order of magnitude. At f), the field is almost completely gone. The latitudinal ionization rate profiles are now completely flat at all heights and at the same level as the non-changing polar regions. Not depicted in Figure \ref{fig:ionization_profiles} are the orderly growth of the dipole components towards pre-reversal magnitude, causing a symmetric and smooth decrease of the central latitude ionization back to the pre-reversal shape.
In the upper panels of Figure \ref{fig:ionization_profiles_vertical}, vertical ionization rate profiles are seen at the first 6 points a) to f) in time during the reversal event at the Equator, 45$^{\circ}$N and for the North Pole. As in the latitudinal profiles, the primary influence of the reversal is seen at central latitudes and increases with altitude. At the poles, the change in ionization is negligible except for at time d), where the magnetic field flares up for a relatively brief period, also affecting the polar regions.

\subsection{Ionization, reversal and solar activity}\label{sec:comparisons_with_sun}
Simultaneous with changes in ionization due to the slowly varying geomagnetic field, changes in solar activity over the 11-year cycle may further modulate the ionization. Therefore, it is instructive to compare the effects of geomagnetic changes and solar activity on atmospheric ionization rates. Solar activity is varied via the solar modulation parameter $\phi$ and can be changed simultaneously with the geomagnetic field. Starting at solar minimum for the pre-reversal magnetic field at time a), the right panel of Figure \ref{fig:ionization_profiles} shows how the ionization rate changes with increasing $\phi$ towards solar-maximum. As the strength of the heliomagnetic field increases, the cosmic ray spectrum diminishes at the lower energies, which reduces the atmospheric ionization rate over the entire globe. However, the effect is strongest at higher altitudes and in the polar regions but weaker around the equator. Note that the change in ionization at high latitudes caused by solar activity is comparable to the change in ionization at the equator caused by the geomagnetic reversal. This tendency is also reflected in the bottom panels of Figure \ref{fig:ionization_profiles_vertical}, which show the change in the vertical ionization profile between solar minimum and maximum for a pre-reversal magnetic field. Here, the change in ionization rate is largest in polar regions and much smaller in the equatorial regions. A contrast to the above is seen in the upper panels of Figure \ref{fig:ionization_profiles_vertical}, where the geomagnetic field varies. Here the geomagnetic field modulates ionization in the equatorial regions much more than in the polar regions, with a magnitude comparable to the changes in the polar areas caused by solar activity.

\begin{figure}\hspace{-1.5cm}\centering
\noindent\includegraphics[width=1.1\textwidth]{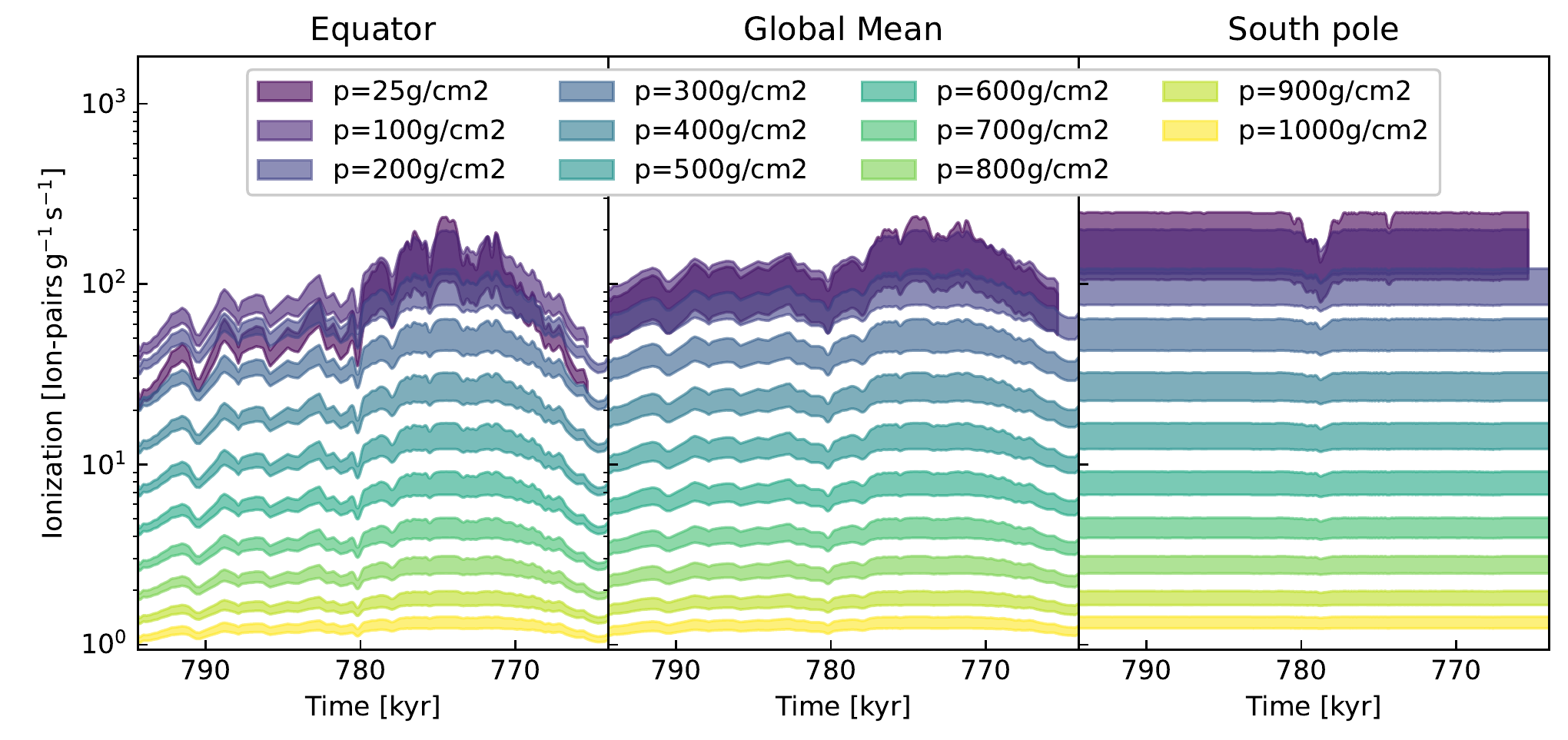}
\caption{Temporal evolution of averaged ion-production rates at 12 different pressure surfaces for the entire span of the magnetic data during the Matuyama-Brunhes reversal event. The average ionization rate is shown between $\pm10$ degrees latitude (left hand panel), for the full globe (center panel), and for 85$^{\circ}$S (right hand panel). At each pressure surface, the upper curve shows the ion-production rate for solar minimum conditions ($\phi=0.3$\,GV), while the lower curve is for the solar maximum conditions ($\phi=1.1$\,GV).}
\label{fig:ionization_temporal}
\end{figure}

The timescales of changes in solar activity and the geomagnetic field reversal are vastly different. Solar activity changes the ionization between solar minimum and maximum on top of the slowly evolving reversal.
To illustrate what that means for ionization at different heights in the atmosphere, we produce ionization rate time series for the full set of magnetic field data form \cite{leonhardt2007}. With 319 available sets of coefficients estimated between 794.47\,kyr and 763.77\,kyr before present, $q(\lambda,\phi,h,t)$ is calculated for each set during the entire period. We calculate the full ionization function for all magnetic fields assuming both solar minimum and maximum conditions. Figure \ref{fig:ionization_temporal} shows this time series for the equatorial region (left panel), the south pole (right panel) and a global mean (middle panel). Each colour corresponds to the ionization at a given atmospheric pressure, and its thickness corresponds to the difference between solar minimum and maximum modulation potential. Again clearly, the biggest change in ionization due to the reversal occurs in the upper atmosphere over the equator, and the lower polar regions are virtually unaffected. Notably, ionization is more sensitive to solar modulation when the geomagnetic field is weakest relative to pre-reversal and post-reversal conditions. Here solar modulation has a minimal effect on ionization rates. 

The amplitude of the solar modulation, i.e.\ the thickness of the bands in Figure \ref{fig:ionization_temporal} depends on the overall strength of the magnetic field. For weaker magnetic fields and thus higher ionization rates, the solar modulation band also thickens. This is better illustrated in Figure \ref{fig:solaramp_vs_time}, which shows the thickness of each band in Figure \ref{fig:ionization_temporal}. In other words, Figure \ref{fig:solaramp_vs_time} shows the amplitude of the solar modulation of atmospheric ionization rates at different heights in the atmosphere throughout the reversal. At the equator, solar modulation increases by more than ten at the top of the atmosphere and 20\% at sea level (CHECK). The dipole dominates Earth's magnetic field and is the most significantly changing part of the field during the reversal. Therefore we plot, in Figure \ref{fig:solaramp_vs_d} the amplitude of the solar modulation of atmospheric ionization as a function of the energy of the dipole component of the field (the Mauersberger–Lowes dipole component). 
Here, the dots show data from the entire time series except between time c) and e) (see Figure \ref{fig:cutoff_rigidity_reversal}) where the dipole component is dominant. Conversely, the transparent x symbols show data from inside this time interval, where the dipole component and the sum of higher-order non-dipole components are of a comparable magnitude. The full lines are a Locally Weighted Scatterplot Smoothing (LOWESS) curve at each atmospheric pressure, excluding the data represented by the x symbols. In this representation, it is clear that the amplitude of the solar 11-year cycle is highly dependent on, and well explained by, the strength of the dipole field, especially in non-polar regions.  

\section{Discussion and conclusion}\label{sec:discussion_conclusion}
\begin{figure}\centering
\noindent\includegraphics[width=1.0\textwidth]{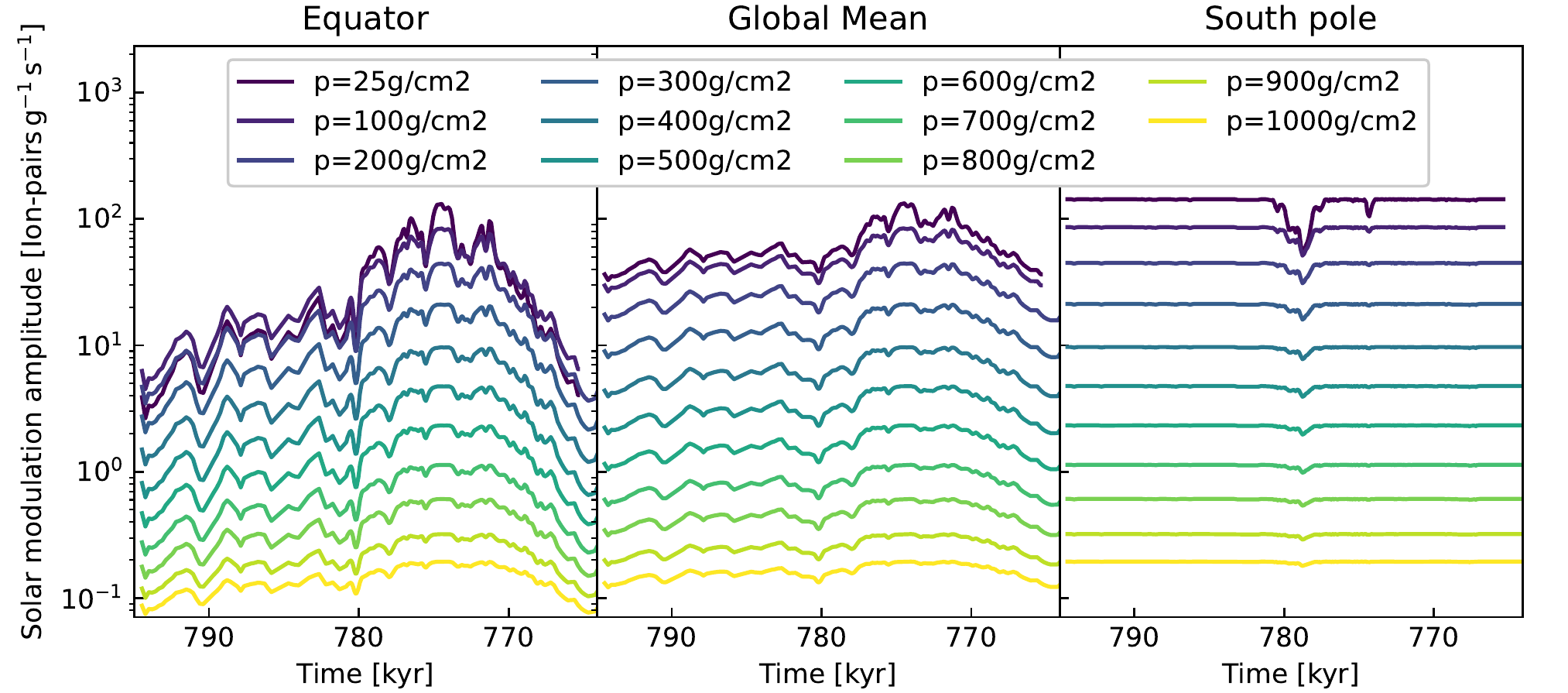}
\caption{Amplitude of the solar modulation of the ionization during the geomagnetic reversal i.e. difference between the ionization with a solar modulation parameter at solar minimum ($\phi=300\,$MV) and maximum ($\phi=1100\,$MV). Colors correspond to different atmospheric pressures, and the three panels show this for the equator (left), the whole globe (middle) and South Pole (right).}
\label{fig:solaramp_vs_time}
\end{figure}

\begin{figure}\centering
\noindent\includegraphics[width=1.0\textwidth]{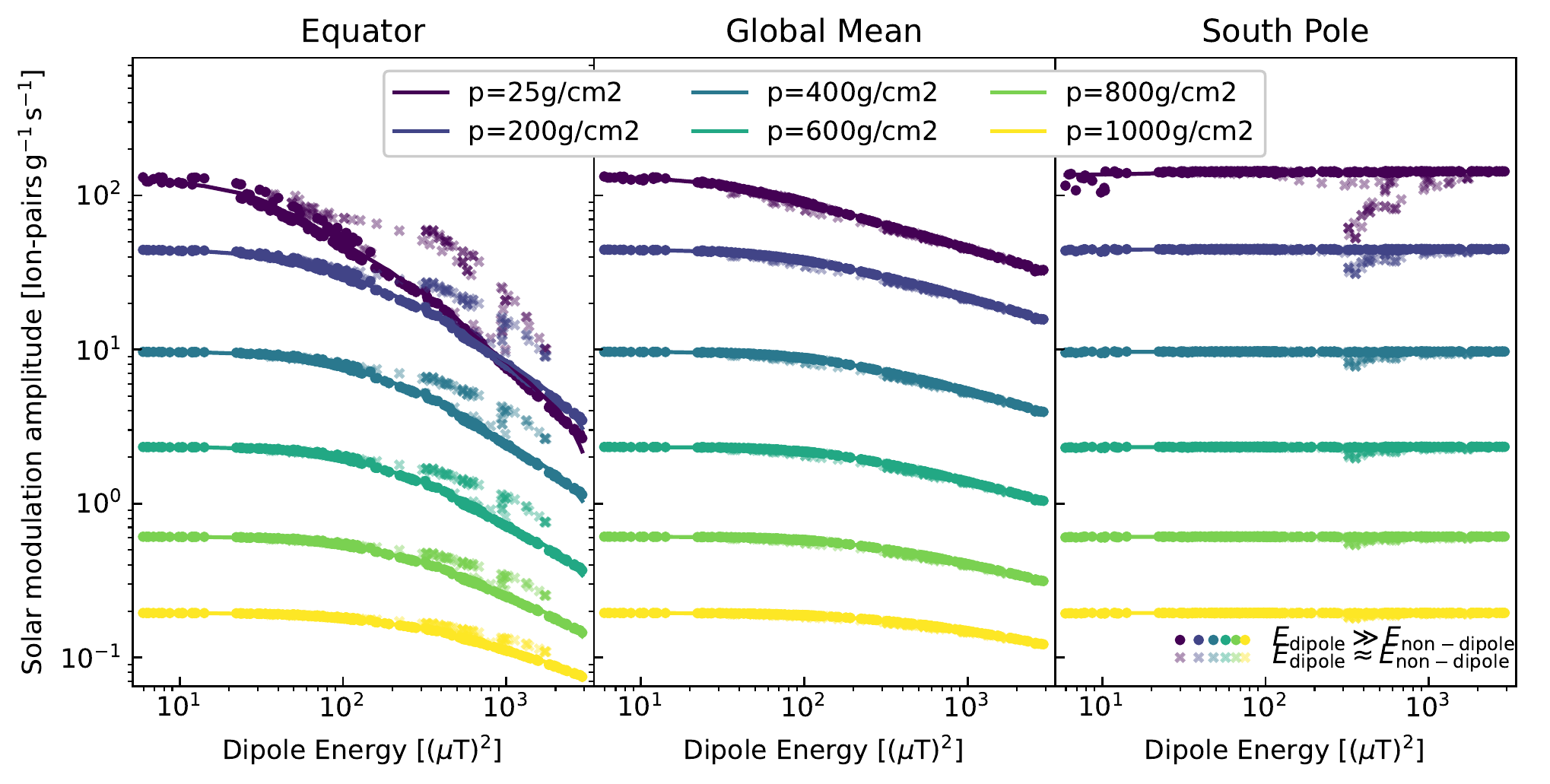}
\caption{Amplitude of the solar modulation as a function on the energy of the dipole component of the Mauersberger–Lowes spectrum for 6 different atmospheric pressures. The opaque dots show data outside the time interval c) to e) (see Figure \ref{fig:cutoff_rigidity_reversal}), as these are times where the multipole is dominant. The transparent x's are the rest of the timeseries, i.e. between time c) and e), where the higher order non-dipole components in total are of a comparable magnitude to the dipole component. The three panels show this for the equator (left), the whole globe (middle) and South Pole (right).}
\label{fig:solaramp_vs_d}
\end{figure}

Due to the vast size of the heliosphere, one might expect a geomagnetic reversal to have less impact on atmospheric ionization than the corresponding reversal of the solar magnetic field through its 11-year cycle. Indeed, how the 11-year cycle affects atmospheric ion concentrations is well-established and known from observations and model studies such as the one included in this study. However, as is apparent from Figures \ref{fig:ionization_profiles}, \ref{fig:ionization_profiles_vertical} and \ref{fig:ionization_temporal}, a geomagnetic reversal affects the ionization in a completely different way, with a magnitude that exceeds that of the solar cycle. Comparing the left and right panels of Figure \ref{fig:ionization_profiles}, it is clear that whereas the solar cycle affects ionization to a large extent globally, the geomagnetic reversal event has its primary effect on non-polar regions. At the equator, the ionization rate increases as the dipole weakens, between a factor of 1.25 and 10, depending on altitude. 
In contrast, the changes in ionization are almost zero in the polar regions during the reversal. One exception is a relatively brief period of about 3000 years right before 780$\,$kyr BP, where the ionization is affected at polar high altitudes. Here is where the influence of the solar 11-year cycle is most prominent. 
Interestingly modulation of the ionization caused by the 11-year solar cycle is highly dependent on the geomagnetic field. Figure \ref{fig:solaramp_vs_time} shows clearly that the solar modulation amplitude increases up to an order of magnitude in the upper atmosphere over the equator during the reversal. Figure \ref{fig:solaramp_vs_d} supports this conclusion by showing the amplitude of the solar modulation as a function of the strength of the dipole. Globally, the difference in ionization between solar-minimum and solar-maximum can change by almost a factor of two or more throughout the atmosphere through a suppressed dipole field. The equatorial regions show an even stronger response, up to nearly two orders of magnitude at the top of the atmosphere.

While the model presented here and similar previous models can reasonably reproduce the observed atmospheric ionization levels, there are no proxies for atmospheric ionization during the reversal. Therefore, modelled ionization levels depend on the accuracy of the paleomagnetic data. Furthermore, the present work relies on the Standard US atmosphere parametrization. Thus, $q(t)$ does not consider paleo-atmospheres. Changes in the atmospheric pressure profile could impact the yield function. Furthermore, modelled solar variations add another modulation of the cosmic ray spectrum to the present model, both through its 11-year cycle and possible long term changes. \citet{usoskin2006} found a $\sim 20\%$ variation in polar ionization rates during the 11-year solar cycle at $p=700\,$g/cm$^2$, which is comparable to the variation seen at a similar height during the reversal. Finally, we assume a cosmic ray spectrum similar to the present-day local interstellar spectrum. While this is not unreasonable, local supernova activity could affect the cosmic ray flux. Therefore, the ionization rates presented here should not be taken as exact but instead as a plausible scenario assuming present-day atmospheric, solar and cosmic spectral conditions. However, the qualitative behaviour of the ionization rate during the reversal still holds: A weaker geomagnetic dipole field causes a rise in ionization. During a geomagnetic reversal, this rise is comparable to the change in polar ionization during the 11-year cycle but instead in non-polar regions. Furthermore, the magnitude of the 11-year cycle is highly dependent on the dipole field as well, so with a weaker field, the resulting ionization would be higher and fluctuate more with solar activity.

The conclusions of the present paper are as follows:
\begin{itemize}
    \item The changes in the geomagnetic field during the Matuyama-Brunhes reversal impact the geomagnetic cutoff rigidity of Earth globally.
    \item This affects the global ionization rates from cosmic rays, which increase as the geomagnetic field weakens.
    \item The effect is particularly prominent in the equatorial region and at all heights of the atmosphere, whereas the polar regions are nearly unaffected.
    \item Ionization levels increase at the equator between 25\% at the surface up to a factor of 6 or more at the top of the atmosphere during the reversal. This corresponds globally to a 13\% ionization change at the sea surface level and $\sim$ 2 at the top of the atmosphere.
    \item The impact of the 11-year solar cycle on ionization depends mainly on the geomagnetic dipole field. During the reversal, the solar modulated ionization increases between a factor of $\sim$2 at the sea surface and almost two orders of magnitude at high altitudes. The solar modulation of atmospheric ionization is most notable at the equator and negligible in the polar regions but still significant when averaged globally. 
\end{itemize}

If atmospheric ions indeed affect terrestrial climate, all available historic geomagnetic variations are of interest. We note that paleomagnetic data are available for other periods, for instance, for the Laschamb excursion \cite{leonhardt2009}, which could provide the basis for future studies of atmospheric ionization and its consequences.

\acknowledgments
All data used in this study are publicly available and referenced in the text. The calculated rigidity maps and the global ionization fields will be made available at an online archive upon publication. 
JS is supported by the Carsberg Foundation. None of the authors or their affiliations have any financial conflicts of interest in relation to this work. 


%
%

\bibliography{agusample.bib}

%
%
%
%
%

\end{document}